\documentclass[showpacs,showkeys,prl,twocolumn,aps]{revtex4}

\newcommand{\ben}{\begin{equation}}
\newcommand{\bens}{\begin{eqnarray}}
\newcommand{\een}{\end{equation}}
\newcommand{\eens}{\end{eqnarray}}

\usepackage{epsfig}

\begin{document}

\title{Phase-field study for the splitting mechanism of coherent misfitting precipitates in anisotropic elastic media}

\author{Pil-Ryung Cha}
\email{dream@plaza1.snu.ac.kr}
\author{Dong-Hee Yeon}
\author{Jong-Kyu Yoon}
%
%\homepage{}
\affiliation{School of Materials Science and Engineering, Seoul National University, Seoul, Korea 151-742.}

\date{\today}

\begin{abstract}
Coherent misfitting precipitates in elastically stressed media such as $\gamma \prime$ particles in nickel-based super-alloys show various
splitting patterns such as doublets, quartets, or octets due to their misfit strain energy. While it is an interesting instability phenomenon
defying conventional surface thermodynamics, its mechanism is not completely clear. Through a phase-field study upon the splitting behavior and
morphological evolution of coherent precipitates, we show that an interface instability driven by elastic anisotropy and a diffusion field can
generate elastically induced splitting during diffusional phase transition. Particle splitting is triggered by interface grooving which advances
by penetrating grooves into the interior of the particle. The sequential evolution of shapes during the splitting process is in good agreement
with previous experiments.

\end{abstract}
%-------------------------------------------------------------------------
%-------------------------------------------------------------------------
%-------------------------------------------------------------------------
\maketitle

% introduction%%%%%%%%%%%%%%%%%%%%%%%%%%%
Elastic stresses arising during solid state phase transitions in some alloys can lead to microstructural features such as precipitate alignment,
cuboidal precipitate shapes, and particle splitting. A wealth of experimental and theoretical works have reported how the mismatch strain
between the precipitate and the surrounding matrix phase changes the characteristics of the microstructure~\cite{Ardell66, Miyazaki79, Yoo95,
Fratzl99, Miyazaki82, Khachaturyan88, Wang93, Zhang98, Lee, Doi92}.

One of the primary reasons for considering the role of elasticity in microstructure formation is that it offers us the possibility of using
elastic stresses to design desirable microstructures. In this context, one of the possibilities often cited is that of inverse coarsening. In
inverse coarsening, the elastic fields promote the growth of a small precipitate at the expense of a larger neighbor - exactly the opposite to
surface energy driven coarsening. In principle, elasticity could favor an array of equally sized precipitates, which would lead to the improved
mechanical properties of the alloy.

Particle splitting is perhaps the most closely linked elastic effect to inverse coarsening. Particle splitting is a process in which one
cuboidal particle splits into several particles, usually two (a doublet) or eight (an octet). Such splitting has been observed
experimentally~\cite{Miyazaki79, Yoo95} and has been theoretically identified as an elastically driven process~\cite{Miyazaki82,
Khachaturyan88}.

While splitting may be an important factor for inverse coarsening, its mechanism and the precise experimental and theoretical conditions under
which it occurs are not completely clear. Some experimental pictures suggest that splitting occurs via a morphological instability in which the
sides of a cuboidal precipitate become increasingly concave prior to splitting~\cite{Yoo95}. Other pictures suggest a hollowing at the center of
the particle~\cite{Miyazaki82}. Many theoretical simulations of splitting have used diffuse interface techniques, and suggest that the
instability occurs by the 'hollowing' mechanism~\cite{Wang93, Zhang98}. However, the hollowing mechanism cannot explain the formation of an
octet and the fact that the splitting occurs mostly in the concave particle shape~\cite{Miyazaki79, Yoo95} though it can describe splitting to
a doublet. Recently, Lee, using a discrete atom method, observed splitting by an instability mechanism~\cite{Lee}.

In this letter we demonstrate, through phase field model calculations, that interface instability driven by elastic anisotropy and diffusion
field can generate particle splitting. Splitting occurred only under the concave growth condition and the picture that a concentrated elastic field around interface groove promotes splitting is very similar to Asaro-Tiller-Grinfeld (ATG) instability~\cite{atg}. In addition, this splitting
mechanism can explain the formation of both doublets and octets.

% Model %%%%%%%%%%%%%%%%%%%%%%%%%%%%%%%%%%%%%
The model in this study assumes coherent interfaces (i.e. there are no dislocations) and takes into account the effect of anisotropic elasticity
and elastic inhomogeneity which represents that the elastic constants of the matrix and the precipitate differ. We assume that the misfit
between the matrix and the precipitate is purely dilatational with magnitude $e^{T}$, the elastic constants of both phases have cubic
(four-fold) symmetry, and that the lattice parameter is not a function of composition. The model is formulated in terms of the concentration
field $c(\vec{r}, t)$, the phase field $\phi(\vec{r}, t)$ which is zero in the matrix phase and one in the precipitate, and the elastic strain
tensor $u_{ij} = 1/2 \left( \partial u_{i} / \partial x_{j} + \partial u_{j} / \partial x_{i} \right)$, where $\vec{u}(\vec{r}, t)$ is the
atomic displacement field. The interfacial region is defined to be a mixture of matrix and precipitate with a different composition but with the
same chemical potential~\cite{skkim}. Therefore, the concentration field of the system is defined as $c=c^{P} h(\phi) + c^{M} [1-h(\phi)]$,
where $c^{A}$ is a concentration in the A-phase and $h(\phi)$ is the smooth interpolation function, which is zero at $\phi=0$ and one at
$\phi=1$ and changes smoothly from zero to one, i.e. $0<\phi<1$. The superscript $P$ and $M$ represent the precipitate and matrix phases,
respectively, which will be used later without description. From the definition of the interfacial region, the following constraint should be
satisfied at any point in the system:
\ben f^{P}_{c^{P}} (c^{P}) = f^{M}_{c^{M}} ( c^{M})
\label{constrnt} \een
where $f^{A}$ is the chemical free energy of the A-phase and $f_{c^A}^A$ represents the derivative of $f^A$ with respect to $c^A$. The free energy
functional is:
\ben {\cal F} = \int_{\vec{r}} \left[ f(\phi, c, u_{ij}) +
{\varepsilon^{2} \over 2} | \nabla \phi |^{2} \right] d\vec{r}
\label{functional} \een
where $f(\phi, c, u_{ij})$ is the free energy density of the system including the coherent strain energy, defined as $w g (\phi) + f^{P} (c^{P})
h ( \phi) + f^{M} (c^{M}) [1- h(\phi)] + f^{el} (\phi, u_{ij})$, where $g(\phi)$ is the double well potential, which has minima at 1 and 0, the
equilibrium values of the precipitates and matrix phases,respectively. In this study, $g(\phi)$ was selected as $\phi^{2} (1-\phi)^{2}$.
$f^{el}$ is the coherent elastic energy density caused by the lattice mismatch between two phases. In general, the elastic energy density is
expressed as \cite{Leo}:
\ben f^{el} = {1 \over 2 } C_{ijkl} (\phi) [u_{ij} - e^{T}
\delta_{ij} \psi (\phi) ] [ u_{kl} - e^{T} \delta_{kl} \psi (\phi)
] \label{elasfunc} \een
where $C_{ijkl}$ is the stiffness tensor that depends on the phase field due to elastic inhomogeneity and $\psi (\phi)$ is a smooth function
with the same characteristics as $h(\phi)$. A summation convention over repeated indices is implicit. Considering the spatial inhomogeneity of
the elastic stiffness, the stiffness tensor is given by:
\ben C_{ijkl} (\phi) = C^{0}_{ijkl} + C^{\prime}_{ijkl} \xi(\phi) \een
where $\xi(\phi)$ is a smooth function which is zero at $\phi=1$ and one at $\phi=0$, $C^{0}_{ijkl} = C^{P}_{ijkl}$, and $C^{\prime}_{ijkl} =
C^{M}_{ijkl}- C^{P}_{ijkl}$, where $C^{A}_{ijkl}$ is the elastic stiffness of the A-phase.

It is reasonable to suppose that the elastic field relaxes much faster than $c$ or $\phi$. The elastic field can then be solved in terms of the
phase field using the condition of local mechanical equilibrium:
\ben {\delta {\cal F} \over \delta u_{i}} = \nabla_{j} \sigma_{ij}
= 0 \label{mechequil} \een
where $\sigma_{ij} = \delta {\cal F} / \delta u_{ij}$ is the stress tensor:
\ben \sigma_{ij} = C_{ijkl} (\phi) [ u_{kl} - e^{T} \delta_{kl} \psi(\phi)] \label{expstress} \een

In this work, in order to obtain accurate solutions of equation (\ref{mechequil}) with minimal calculation time, the high order approximation
method similar to the method proposed by Chen and coworkers~\cite{Chen} is used instead of the iteration method which gives very accurate
solutions but requires considerable calculation time.

The solution of equation (\ref{mechequil}) to zeroth order in the elastic stiffness
$C^{\prime}_{ijkl}$ is
\ben \hat{u}^{0}_{k} = i \alpha^{-1}_{ki} \sigma^{0}_{ij} q_{j} \hat{\psi} (\hat{\phi}) \label{zerothsol} \een
where $\hat{A}$ represents the value of $A$ in Fourier space, $q_{j}$ the \emph{j}-direction component of the wave number vector,
$\sigma^{0}_{ij} = C^{0}_{ijkl} e^{T} \delta_{kl}$, and $\alpha^{-1}_{ki}$, indicates a component of the inverse tensor of $\alpha_{ik}$ which
is defined as $C^{0}_{ijkl} q_{j} q_{l}$. Then, the displacement field to the \emph{n}th order becomes
\bens \hat{u}^{n}_{k} = \hat{u}^{0}_{k} & + & i \alpha^{-1}_{ki} C^{\prime}_{ijlm} e^{T} \delta_{lm} q_{j} \{
\xi(\phi) \psi(\phi)\}_{F} \nonumber \\
& - & i \alpha^{-1}_{ki} C^{\prime}_{ijlm} q_{j} \{ [\xi(\phi) \nabla_{m}] u^{n-1}_{l} \}_{F} \label{nthsol} \eens
where $u^{n-1}_{k}$ is the displacement field to the \emph{n-1}th order and $\{ \}_{F}$ represent the Fourier transform of the term in braces.

The elastic field can now be expressed in terms of the phase field $\phi$. Substituting the solution for the strain field gives the free energy
in terms of $c$ and $\phi$. Assuming that the system evolves in time so that its total free energy decreases monotonically, the evolution
equation for the phase field becomes:
\ben {\partial \phi \over \partial t} = - M {\delta {\cal F} \over \delta \phi} = M \left(
\varepsilon^{2} \nabla^{2} \phi - f_{\phi} \right), \label{governphi} \een
with
\ben f_{\phi} = h' (\phi) \{ f^{P} - f^{M} + (c^{M} - c^{P}) f^{M}_{c^{M}} \} + f^{el}_{\phi} + w
g'(\phi) \label{fphi} \een
while the concentration field should evolve with:
\bens {\partial c_{k} \over \partial t} & = & \nabla \cdot {D(\phi) \over f_{cc}} \nabla {\delta
{\cal F} \over \delta
c} \nonumber \\
& = & \nabla \cdot D(\phi) \nabla c + \nabla \cdot {D(\phi) \over f_{cc} } f_{c\phi} \nabla \phi
\label{governconc} \eens
where $M$ is the phase-field mobility, $D(\phi)$ is the diffusivity of solute atoms, and $f_{\phi}$, $f^{el}_{\phi}$, $f_{cc}$, and $f_{c\phi}$
are $\partial f /
\partial \phi$, $\partial f^{el} / \partial \phi$, $\partial^{2} f / \partial c^{2} $, and $\partial^{2} f / \partial c \partial \phi$,
respectively.

In the sharp interface limit, the phase field model presented here reduces to a modified Gibbs-Thompson equation with a coherent interface as
follows:
\ben \alpha \sigma v_{n} = \kappa_{c} \sigma + \Delta \mu_{C} + G^{coh} + G^{trans} \label{GTeq}
\een
Here, $\alpha = D / M \varepsilon^{2}$, $v_{n}$ is the interface velocity, $\sigma$ the interface energy, $\kappa_{c}$ the curvature of the
interface, and $\Delta \mu_{C}$ the chemical driving force of the phase transition defined as $\mu^{P}_{C} - \mu^{M}_{C}$, where $\mu^{A}_{C}$
represents the chemical potential at the A-phase side of the interface. $G^{coh}$ means the elastic energy required to maintain coherence at the
interface, defined by $(u^{M}_{ij} - u^{P}_{ij})\sigma^{M}_{ij}$, where $u^{A}_{ij}$ and $\sigma^{A}_{ij}$ are elastic strain and stress tensors
at the A-phase side of the interface, respectively~\cite{LeoGT}. $G^{trans}$ represents the difference in the elastic strain energies of the
precipitate and the matrix, as defined by $1/2 \sigma^{P}_{ij} ( u^{P}_{ij} - e^{T} \delta_{ij}) - 1/2 \sigma^{M}_{ij} u^{M}_{ij}
$~\cite{LeoGT}. The detailed derivation of Eq. (\ref{GTeq}) will emerge elsewhere~\cite{further}.

% Result and Discussion %%%%%%%%%%%%%%%%%%%%%%%%%%%%%%%%%%%%%%%%%%%
Numerical simulations on a discrete lattice were performed in two-dimensions. Euler's method was used for the integration in time. For all
simulations presented here, the mesh size $\triangle x = 1.0\times 10^{-8} m$, the time step $\triangle t = 0.0125$, and $\varepsilon = 3.52
\times 10^{-5}$. This choice of $\triangle x$ and $\varepsilon$ guarantees that the interface is resolved by at least five points and the
interfacial energy is 0.0243 $J/m^2$. We assume that the diffusivities of two phases are the same and $1 \times 10^{-15} m^{2}/s$. Quadratic functions were used for the free energy functions of both phases for simplicity, that is, $f^{P}
= \Delta^{P} (c^{P} - c^{P}_{e} )^2 / V_{m}$ and $f^{M} = \Delta^{M} (c^{M} - c^{M}_{e} )^2 / V_{m}$, where $\Delta^{P}$ and $\Delta^{M}$ were
10000 and 2000, respectively, $V_m$ is molar volume and was set at $7.0 \times 10^{-6} m^{3}/ mol$, and $c^{P}_{e}$ and $c^{M}_{e}$ are the
equilibrium concentrations without elastic field and were set at 0.2 and 0.4, respectively. The elastic constants of two phases (in units of
$10^{11} erg/cm^3$) are $C^{P}_{11}=1.67$, $C^{P}_{12}=1.07$, $C^{P}_{44}=0.99$, $C^{M}_{11}=1.12$, $C^{M}_{12}=0.63$, and $C^{M}_{44}=0.57$,
respectively; for the Ni-based superalloy~\cite{Miyazaki82}. The misfit strain between the precipitate and the matrix is dilatational and
assumed to be $0.8 \% $. The displacement field was calculated to the third order from Eq. (\ref{nthsol}) because the inhomogeneities of the
elastic constants were very large. Periodic boundary conditions were employed in all directions.

\begin{figure}[htbp]
\centerline{ \epsfig{figure=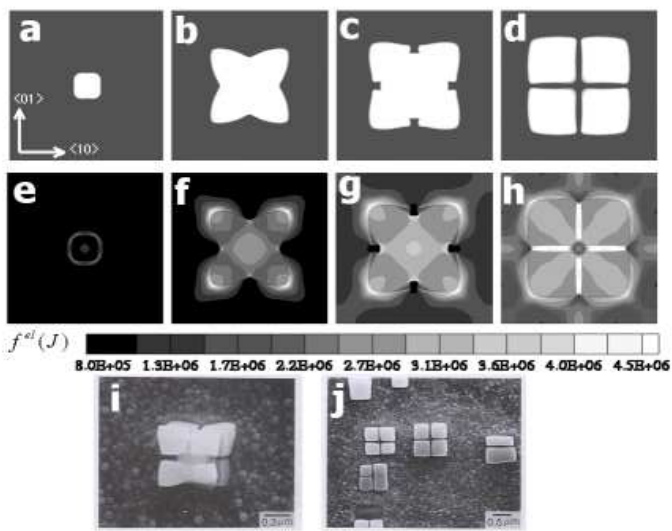, width=3in, height=3.7in, angle=90, silent=}} \caption{Sequential evolutions of the
precipitate (the top row) and the corresponding misfit elastic energy density distributions(the second row): from the left, $t=3.125, 31.25,
62.5$ and $150 s$. The lowest row shows splitting and split $\gamma \prime$-precipitates in the Ni-based superalloy system~\cite{Yoo95}. In the
calculation, $M=5.0 \times 10^{-7}$. The total size of the calculation domain is $2.56 \mu m \times 2.56 \mu m$.} \label{fig1}
\end{figure}

Figure~\ref{fig1} shows the calculated splitting behavior of the precipitate and the corresponding elastic energy density distribution at
various times. The morphological instability of the interface due to the diffusion field and anisotropy of the elastic stiffness is evident and
this interface instability leads to particle splitting. Despite the simplicity of the two-dimensional calculation, the splitting pattern is in
good agreement with the experimental results (see Fig.~\ref{fig1}i and j). The scenario of the splitting is as follows: initially the
precipitate grows in a circular shape due to the isotropic interface energy, and then transforms to cuboidal shape above a critical size,
because the misfit strain energy proportional to its volume dominates the interfacial energy in proportion to the interfacial area. Subsequently,
the growth rate in the $<11>$ direction is enhanced due to a high chemical driving force while the migration rate in the $<10>$ direction is
suppressed due to enriched solute and a corresponding low chemical driving force. Therefore, the growing shape transforms to a concave shape. In
the concave growth regime, the interface in the $<10>$ direction has a negative curvature, and so, the elastic strain energy concentrates around
the concave edge . This concentrated elastic energy drives the dissolution of the precipitate, and then a groove forms and penetrates the
precipitate (see Fig.~\ref{fig1}c). This situation resembles Asaro-Tiller-Grinfeld instability, in which elastic media under the nonhydrostatic
stress show the morphological instability~\cite{atg}. Through various numerical simulations, we confirmed that the splitting occurs only in the
concave growth condition and that it is induced by the interface instability. This is in good agreement with the experimental
observations~\cite{Ardell66, Miyazaki79, Yoo95}.

\begin{figure}[hbtp]
\centerline{ \epsfig{figure=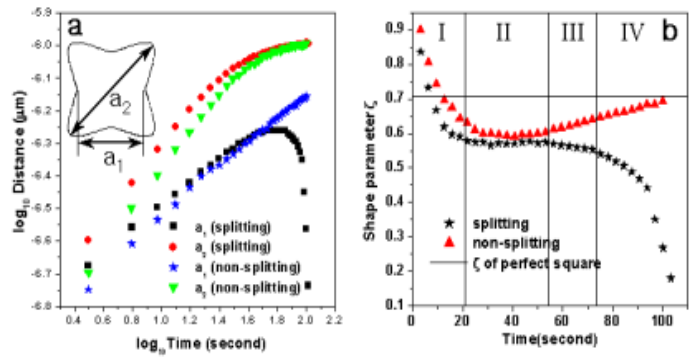, width=2in, height=3.7in, angle=90, silent=}} \caption{Time
variations of $<10>$ length ($a_1$) and $<11>$ length ($a_2$) of the precipitate (a) and
the corresponding ratio of $a_1$ to $a_2$ (b) which is defined as the shape parameter $\zeta$. The inset shows the
definitions of $a_1$ and $a_2$. In the case of splitting, $M=5.0 \times 10^{-7}$ and in non-splitting, $M=3.0
\times 10^{-7}$.} \label{fig2}
\end{figure}

Figure~\ref{fig2} shows the sequential variations of the $<10>$ length ($a_{1}$) and $<11>$ length ($a_{2}$) of the precipitate and the ratio of
the two lengths which is called a shape parameter ($\zeta$) hereafter. There are four different growth modes (see Fig.~\ref{fig2} b). In the
first growth mode the growth rates of $a_{1}$ and $a_{2}$ are proportional to $t^{0.35 \pm 0.005}$ and $t^{0.55 \pm 0.01}$, respectively and
hence transition from circular through cuboidal to concave shapes occurs (for a perfect square $\zeta=0.7$). The second regime can be called the
shape-conserving mode because the shape parameter remains constant in this period. The growth rates of $a_{1}$ and $a_{2}$ have the same
dependence $t^{0.31 \pm 0.01}$ on time. The growth of $a_{1}$ stops upon entering the third growth mode while $a_{2}$ continues to grow and the
shape parameter starts to decrease again. In this regime the grooves are generated on the four interfaces of the $<10>$ direction (see
Fig.~\ref{fig1}c) and concavity around the grooves starts to relax because the elastic field around the grooves is relieved by local
concentration of the stress field on the edges of the grooves. One can see this by comparing Fig.~\ref{fig1}b with Fig.~\ref{fig1}c. Once the
elastic energy concentrated on the tips dominates the chemical driving force for precipitate growth which is reduced due to the buildup of
solute around the grooves, the grooves penetrate into the precipitate and the fourth growth stage begins. Note that the particle in the
non-splitting case also shows concave growth, but then returns to the cuboidal shape with time (see the variations of the filled triangles in
Fig.~\ref{fig2}b).

\begin{figure}[hbtp]
\centerline{ \epsfig{figure=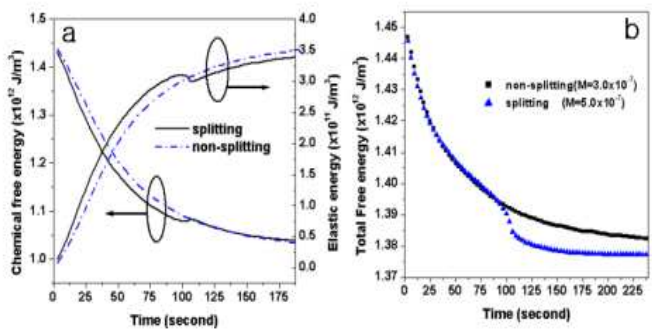, width=2in, height=3.7in, angle=90, silent=}} \caption{Time variations of the total free energy
(Eq.~\ref{functional}), elastic energy (a), and the total free energy, except the elastic energy (b). The phase field mobilities are $5.0 \times
10^{-7}$ for the splitting case and $3.0 \times 10^{-7}$ for the non-splitting case, respectively. } \label{fig3}
\end{figure}

Fig.~\ref{fig3} shows the variations of the total free energy [Eq. (\ref{functional})], elastic energy, and the total free energy excluding the
elastic energy, with the evolution of the second particle. Here, the third free energy term mentioned above is called the chemical free energy
although it includes the interfacial energy (i.e. gradient energy and double well potential energy). During the evolution of the particle, the
total free energy and the chemical free energy decrease while the strain energy increases due to misfitting particle growth. When the splitting
has been almost completed, the chemical free energy jumps due to the increased interfacial area, but the total strain energy reduces by
interface grooving, which actually resembles the ATG instability~\cite{atg}. As the decrease of strain energy overwhelms the increased
interfacial free energy, the total free energy decreases and the splitting process advances. As shown in Fig.~\ref{fig3}b, the configuration of
four split pieces is a more stable state than non-split single particle. Hence the non-split large particle is considered as a meta-stable phase
and the activation barrier is the interfacial free energy corresponding to the increased interfacial area. Considering these points, the
splitting phenomenon is kinetically rather than energetically driven.

In this work, it has been shown through phase field study that the interaction between the elastic field condensed at the concave region and
solute enrichment causes the particle splitting phenomenon. The splitting is mediated by interface instability, which differs from previous
studies that splitting occurs through the nucleation of the matrix phase at the center of the precipitate~\cite{Miyazaki82, Wang93} or that it
is caused by energy considerations~\cite{Khachaturyan88, Doi92}. In particular, the sequential splitting behavior realized by our simulation
shows an excellent agreement with previous experimental studies~\cite{Miyazaki79, Yoo95}.

\end{document}